\documentclass{article}
\usepackage{spconf,amsmath,graphicx}
\usepackage{float}
\usepackage{amsmath}
\usepackage{amssymb}
\usepackage{amsfonts}
\usepackage{enumitem}
\setlist{nosep, leftmargin=14pt}
\usepackage{mwe}

\title{Denoising study of Fluoroscopic Images in real time tumor tracking System based on Statistical model of noise}
\twoauthors{Yongxuan Yan \thanks{This work was supported by JST SPRING, Grant Number JPMJSP2111.}, Fumitake Fujii}
{Graduate School of Sciences and Technology \\ for Innovation, Yamaguchi University}
{Takehiro Shiinoki}{Department of Radiation Oncology, Graduate \\ School of Medicine, Yamaguchi University.}
\begin{document}
\maketitle
\begin{abstract}
This study investigates the noise characteristics of intraoperative X-ray fluoroscopic images acquired during real-time image-guided radiotherapy (IGRT), and presents a novel noise image generation method based on the identified noise amplitude and spatial probability patterns. Initially, noise-free digitally reconstructed radiographs (DRRs) were generated using patient CT data combined with projection algorithms and the spatial configuration of the real-time tumor tracking system. Based on the observed noise probability and amplitude distributions, noise was then added to these DRRs to create Dataset 1. As a control, Dataset 2 was generated by adding Gaussian noise with the same mean and variance as Dataset 1; however, the noise probability in Dataset 2 is independent of pixel location and pixel intensity. Both datasets were used to fine-tune a pre-trained SwinIR model with identical training parameters. Tests on phantom images containing real noise show that the SwinIR model trained with the proposed noise model dataset achieves superior denoising performance over the model trained with Gaussian noise and the model without transfer learning, with an average PSNR improvement of 1.45 dB. This study contributes to a deeper understanding of noise patterns in these fluoroscopic images and is crucial for enhancing image quality and the accuracy of real-time tumor tracking in radiotherapy.
\end{abstract}
\begin{keywords}
Image-guided radiotherapy (IGRT), Fluoroscopic images, Denoise, Noise statistical model
\end{keywords}
\section{Introduction}
\label{sec:intro}
Stereo X-ray tumor tracking systems are widely used in IGRT for lung cancer\cite{yan2024Access}. Investigating noise characteristics in these X-ray fluoroscopic images is essential to improve image quality and the subsequent accuracy of real-time tumor tracking. The X-ray tube is typically embedded beneath the floor while the imaging plane is mounted on the ceiling in stereo X-ray imaging systems. This alignment allows the linear accelerator and treatment table to move freely\cite{cyberknife,shiinoki}. This unique geometry significantly increases the object-to-image distance (OID),

\begin{figure}[H]
\begin{minipage}[b]{1.0\linewidth}
  \centering
  \centerline{\includegraphics[width=8cm]{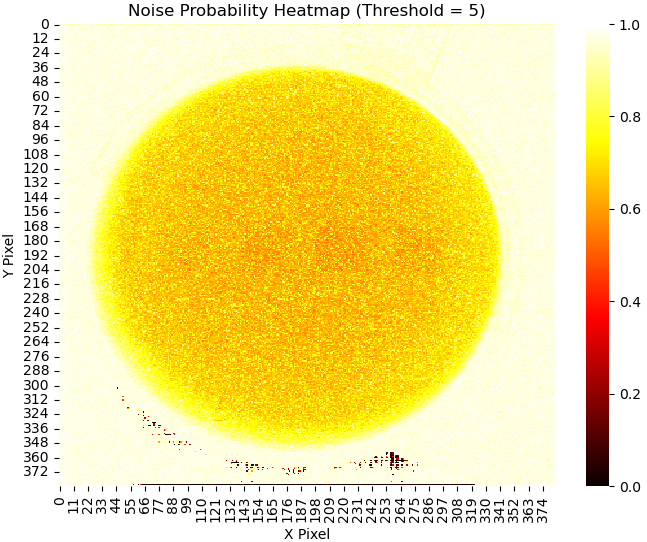}}
  \centerline{(a) Noise probability distribution}
\end{minipage}
\begin{minipage}[b]{.44\linewidth}
  \centering
  \centerline{\includegraphics[width=3.7cm]{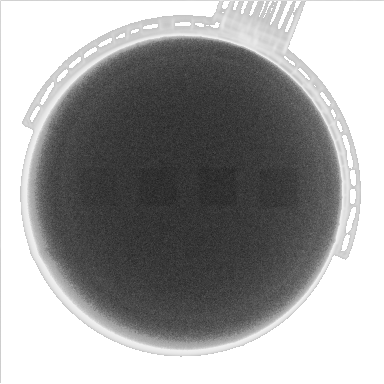}}
  \centerline{(b) Fluoroscopic image}
\end{minipage}
\hfill
\begin{minipage}[b]{0.54\linewidth}
  \centering
  \centerline{\includegraphics[width=4.45cm]{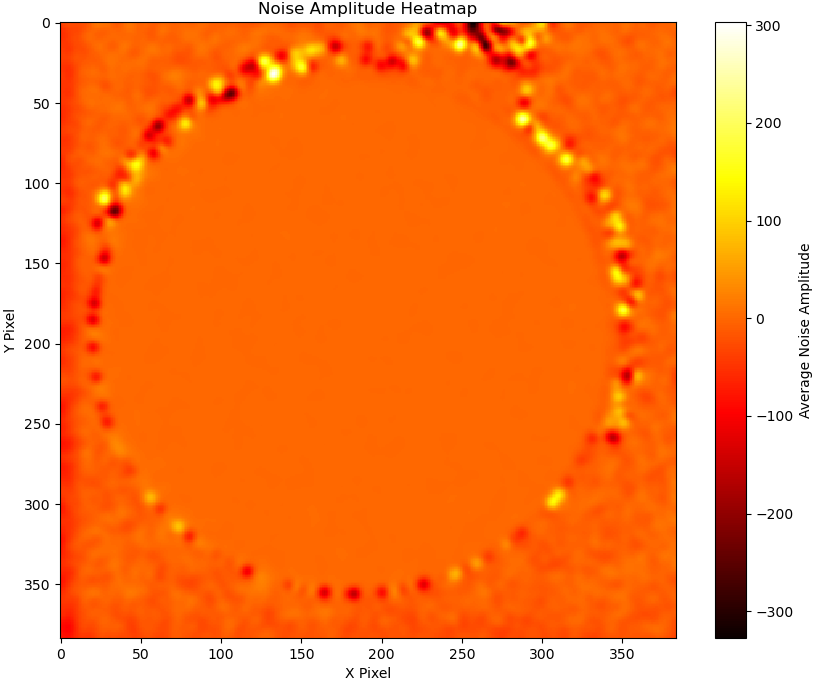}}
  \centerline{(c) Average noise amplitude}
\end{minipage}
\caption{Noise probability distribution and amplitude in a SyncTraX fluoroscopic X-ray image. (a) This heatmap reveals lower noise probabilities in the square regions corresponding to embedded aluminum blocks, while in the circular area representing the gelatin, the noise probability decreases as the distance to the image center decreases. (b) The contrast of this image has been enhanced to clearly show the aluminum blocks. The central circular area represents the gelatin, inside which four aluminum blocks of varying thicknesses are placed to simulate human tissue and bones. (c) This heatmap shows the distribution of average noise amplitude across the entire image. The noise amplitude is lowest and smoothest within the gelatin, while the amplitude peaks and fluctuates sharply at the gelatin boundaries.}
\label{fig.1}
\end{figure}

\noindent  which exacerbates noise in the images. We use a real-time tumor tracking system SyncTraX (Shimadzu, Co. Ltd, Kyoto, Japan) in this study. The OID of this imaging system is 2091 with 10 mm installation tolerances. The increased OID may result in higher noise levels in the captured images, sharply contrasts with traditional diagnostic X-rays where the patient is positioned as close as possible to the imaging plane.

Despite the importance of improving the image quality of such fluoroscopic images for radiotherapy, there is limited research on denoising algorithms specifically designed for these images. Most existing efforts have focused on diagnostic X-ray images\cite{zhenduanX1} or CT images\cite{ct}, leaving a gap in the literature for applications specific to IGRT.

\section{Materials and Methods}
\label{sec:format}
\begin{figure}[ht]
\begin{minipage}[b]{.48\linewidth}
  \centering
  \centerline{\includegraphics[width=3.3cm]{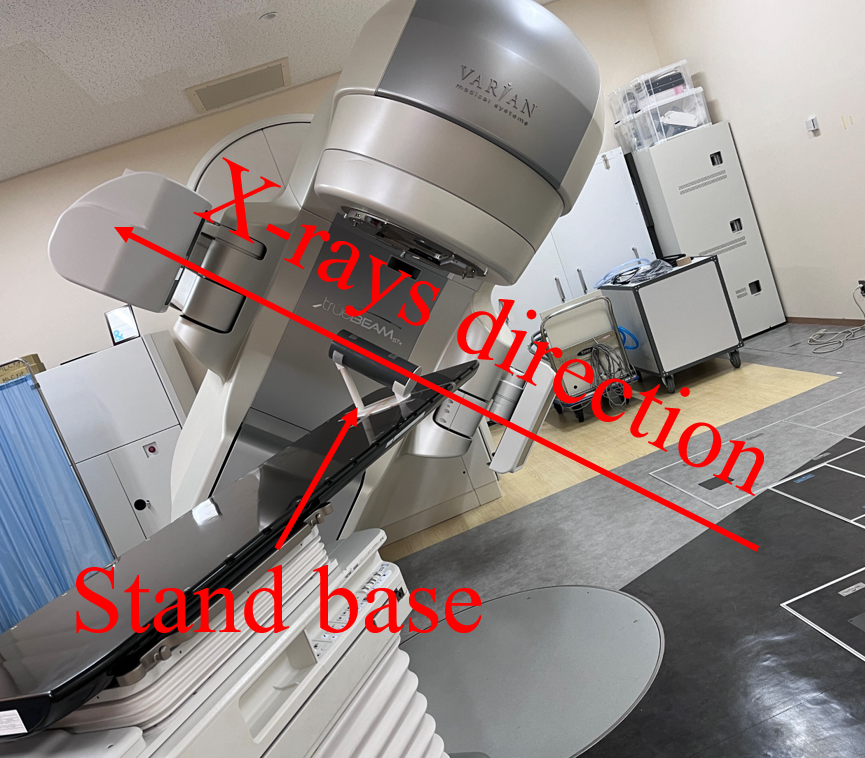}}
\end{minipage}
\begin{minipage}[b]{.48\linewidth}
  \centering
  \centerline{\includegraphics[width=4.5cm]{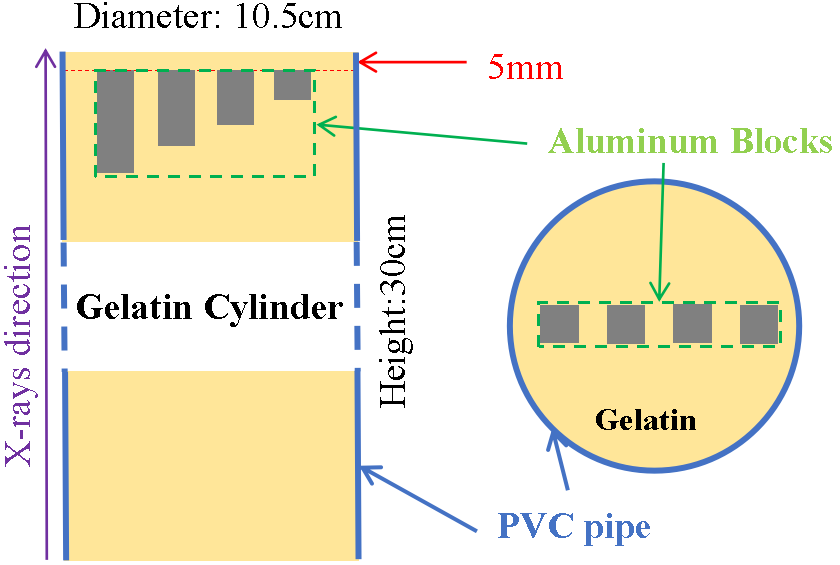}}
\end{minipage}
\caption{Image acquisition site (left) and the gelatin cylinder (right) with aluminum blocks which are numbered as 4, 3, 2, and 1 from left to right.}
\label{fig.2}
\end{figure}

\begin{figure}[ht]
\begin{minipage}[b]{.28\linewidth}
  \centering
  \centerline{\includegraphics[width=2.5cm]{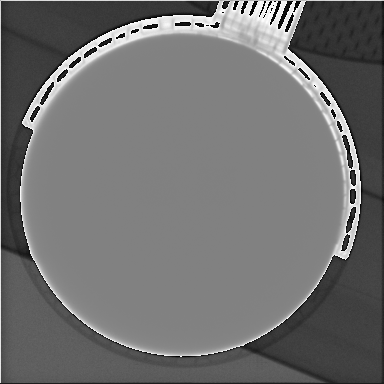}}
\end{minipage}
\begin{minipage}[b]{.31\linewidth}
  \centering
  \centerline{\includegraphics[width=2.5cm]{Global_Org_1.png}}
\end{minipage}
\begin{minipage}[b]{.34\linewidth}
  \centering
  \centerline{\includegraphics[width=2.5cm]{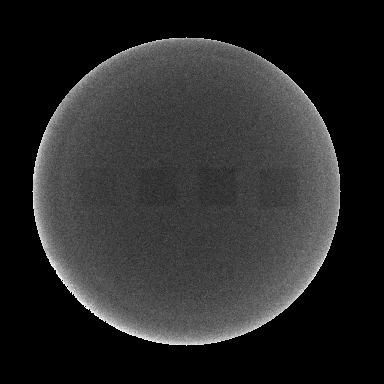}}
\end{minipage}
\caption{Pre-processed images from collected 16-bit raw data. From left to right are (1) original image. In the background, the area with relatively higher brightness in the lower left corner is the air, whereas the rest corresponds to the treatment couch. (2) Contrast-enhanced image, and (3) image with the non-gelatin regions removed.}
\label{fig.3}
\end{figure}

To investigate the noise characteristics in the SyncTraX fluoroscopic images, two cylindrical gelatin phantoms were created: one entirely composed of gelatin, and the other embedded with aluminum blocks as shown in Fig.\ref{fig.2}. 300 X-ray fluoroscopic images were taken under '100KV, 80mA, 4ms' X-ray condition. The acquired 16-bit raw data were preprocessed to obtain uint8 384 × 384 images for imaging group 3, as shown in Fig. \ref{fig.3}.

For each group of images, we calculated the difference between each image and the reference image. The reference images \( I_{\text{ref}} \) were obtained by averaging all captured images of each group

\[
I_{\text{ref}} = \frac{1}{N} \sum_{i=1}^{N} I_i ,
\]
where \( N \) is the total number of images and \( I_i \) is each individual image in a group. We define the noise image \( Noise_i \) as follows:

\[
Noise_i = I_i - I_{\text{ref}}
\]

To calculate the noise probability, a threshold \( T \) is defined, and the noise probability at a pixel \( P(x,y) \) is calculated as the frequency of its value in \( Noise_i \)  exceeding the threshold:

\[
P(x,y) = \frac{1}{N} \sum_{i=1}^{N} \mathbb{f}\left( |Noise_i(x,y)| > T \right)
\]

where the indicator function \( \mathbb{f}(\cdot) \) is defined as follows:

\[
\mathbb{f}( |Noise_i(x,y)| > T ) = 
\begin{cases} 
1, & \text{if } |Noise_i(x,y)| > T \\
0, & \text{if } |Noise_i(x,y)| \leq T
\end{cases}
\]

\begin{figure}[ht]
\begin{minipage}[b]{.28\linewidth}
  \centering
  \centerline{\includegraphics[width=2.5cm]{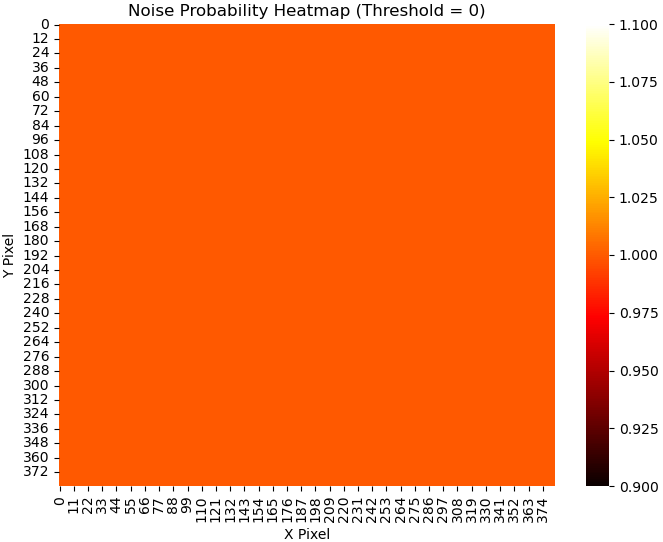}}
  \centerline{(a) Threshold = 0}\medskip
\end{minipage}
\begin{minipage}[b]{.31\linewidth}
  \centering
  \centerline{\includegraphics[width=2.5cm]{noise_heatmap_threshold_5.png}}
  \centerline{(b) Threshold = 5}\medskip
\end{minipage}
\begin{minipage}[b]{.34\linewidth}
  \centering
  \centerline{\includegraphics[width=2.5cm]{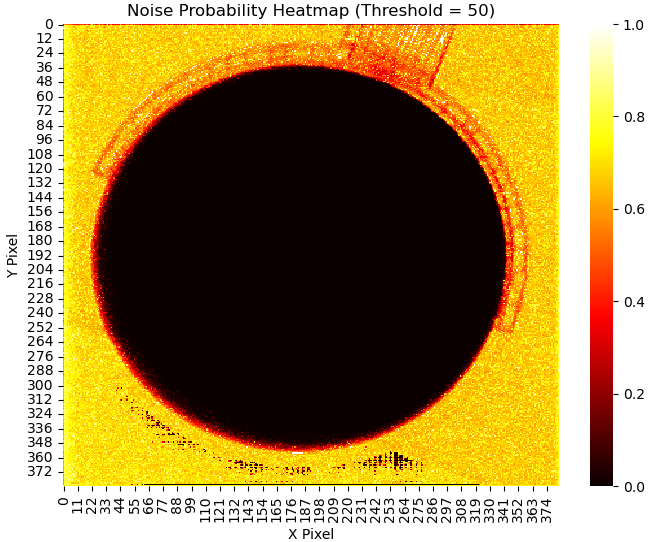}}
  \centerline{(c) Threshold = 50}\medskip
\end{minipage}
\caption{Noise probability heatmaps under different thresholds.}
\label{fig.4}
\end{figure}

This allowed us to isolate the noise from the background. We altered the noise detection threshold T from 0 to 50, with an increment of 1, to evaluate how different thresholds impacted noise probability. Noise probability heatmaps were generated for each threshold as illustrated in Fig.\ref{fig.4}. We also created amplitude distribution maps to visualize noise amplitude across the entire image space as illustrated in Fig.\ref{fig.1} (c). 

These maps provide a comprehensive view of the noise characteristics in the fluoroscopic images. We generated noise onto noise-free DRR images based on the spatial distribution and amplitude characteristics of the observed noise probability. 800 DRR images were generated using a ray-projection algorithm on patient CT data, and both the DRR and noise images were used as training data for transfer learning of the pre-trained SwinIR\cite{SwinIR} model. A control groups were designed for comparison by using a traditional Gaussian noise generation algorithm to create standard noisy images for SwinIR training. We refer to the models as SwinIR-c and SwinIR-n to represent the SwinIR model trained with our noise images, and the model trained with standard Gaussian noise images, respectively. The training batch size is 256, with initial learn rate of 0.001 and 10 epochs.

As the patients move their body based on respiration and heartbeat, it was not feasible to create noise-suppressed reference images through averaging. Therefore, we used phantom images of types (2) and (3) from Fig.\ref{fig.3}  to test the SwinIR models, with reference images as the ground truth for model evaluation.

\section{Experimental Results}
\label{sec:pagestyle}
\begin{figure}[ht]
\begin{minipage}[b]{1\linewidth}
  \centering
  \centerline{\includegraphics[width=7cm]{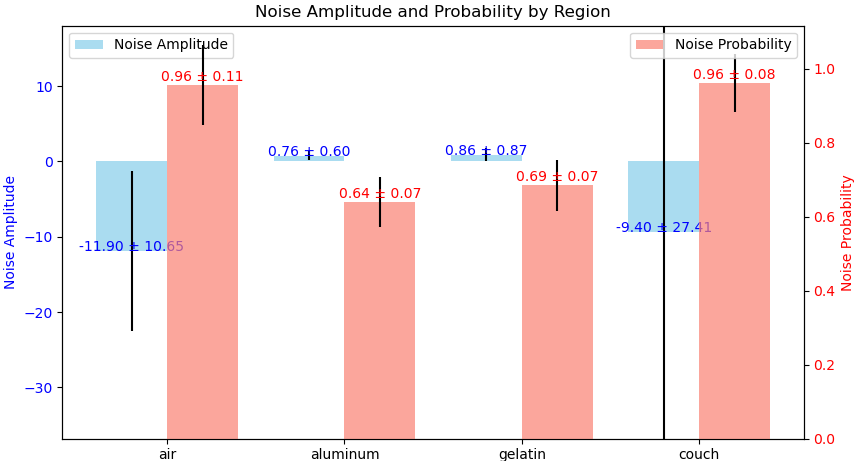}}
\end{minipage}
\caption{Mean and standard deviation of noise in the air, aluminum block, gelatin, and  couch regions. The figure highlights the distinct noise characteristics in each region, with notable differences in noise probability and amplitude distribution.}
\label{fig.5}
\end{figure}

\begin{figure}[ht]
\begin{minipage}[b]{1\linewidth}
  \centering
  \centerline{\includegraphics[width=6.5cm]{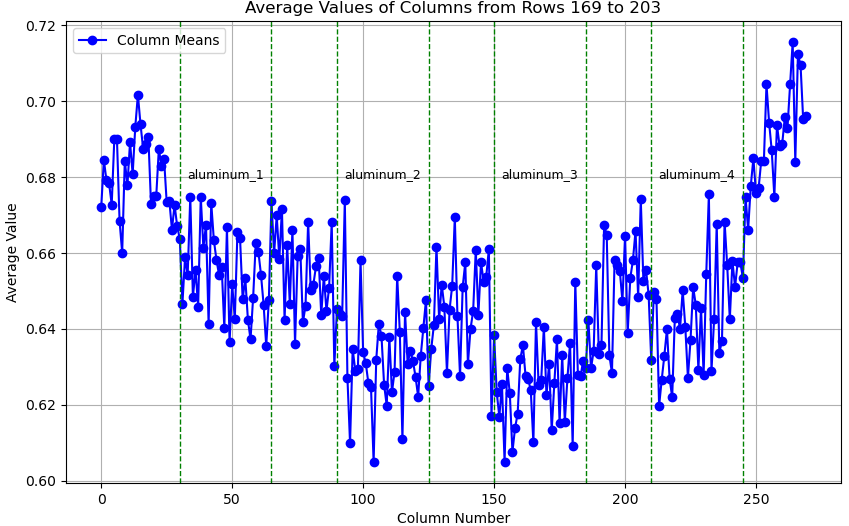}}
\end{minipage}
\begin{minipage}[b]{1\linewidth}
  \centering
  \centerline{\includegraphics[width=6.5cm]{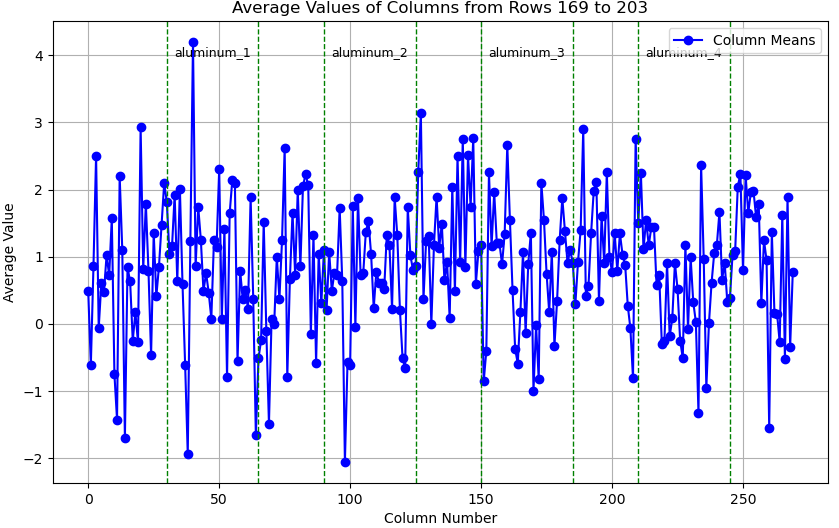}}
\end{minipage}
\caption{The average value change of the aluminum block columns, rows from 169th to 203rd contains aluminum blocks, and pixels at the edge of the image were removed to better observe the changes inside the gelatin. (a) is the probability change while (b) is the amplitude change. It can be seen that the noise probability in the aluminum block columns is lower than that in the gelatin area, while the noise amplitude does not change significantly.}
\label{fig.6}
\end{figure}

 The spatial probability distribution and amplitude characteristics of the noise are summarized in Fig.\ref{fig.5} and Fig.\ref{fig.6}. Based on these statistical models, we developed a method for calculating noise probability. For each pixel in the DRR images generated in this experiment, the composite noise probability is determined by the basic global noise probability shown in Fig.7 and the pixel’s intensity. Lower pixel values indicate greater X-ray absorption at that pixel, leading to a lower noise probability. The composite noise probability \( P(i,j) \) in this experiment is calculated as follows:
\[
P_{(i,j)} = P_{f(i,j)} + V_{(i,j)} \cdot \delta,
\]
where \( P_{f(i,j)} \) is the basic noise probability of pixel \( (i, j) \), \( V_{(i,j)} \) represents the intensity of a pixel at \( (i, j) \), and \(  \delta \) is a constant, approximately \(  \delta \approx 0.001 \) in this experiment.

Since the air and treatment couch areas are not regions of interest and can be easily processed, noise was not generated for these areas. Thus, the noise amplitude is simplified as Gaussian noise with a mean of 5.8 and a standard deviation of 5.8. The noisy images used for training SwinIR-n were generated with the same mean and standard deviation. The results of the noise image generation are demonstrated in Fig.\ref{fig.8}. And the denoise results are illustrated in Fig.\ref{fig.9}. 

The models were evaluated using peak signal to noise ratio (PSNR) and visual information fidelity (VIF) as a performance metric. The results are shown in Tab.\ref{Tab.1}.

\begin{table}[]
\centering
\setlength{\tabcolsep}{6mm}{
\begin{tabular}{lll}
\hline
             & SwinIR-n & SwinIR-c \\ \hline
Avg PSNR(dB) & 29.997   & 31.452   \\
Avg VIF      & 0.267    & 0.367    \\ \hline
\end{tabular}}
\caption{Models evaluation results.}
\label{Tab.1}
\end{table}

\begin{figure}[h]
\begin{minipage}[b]{.48\linewidth}
  \centering
  \centerline{\includegraphics[width=2.5cm]{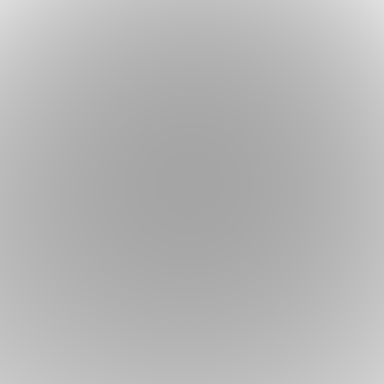}}
\end{minipage}
\begin{minipage}[b]{.48\linewidth}
  \centering
  \centerline{\includegraphics[width=2.5cm]{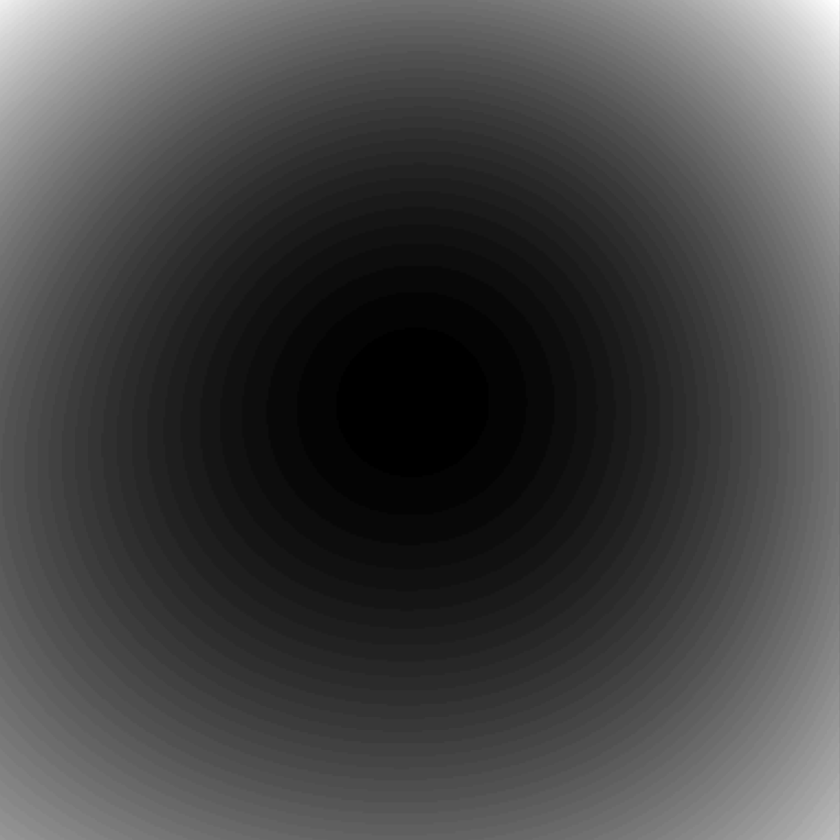}}
\end{minipage}
\caption{Visualization of the basic global noise probability(left) and contrast stretched(right) result.}
\label{fig.7}
\end{figure}

\begin{figure}[h]
\begin{minipage}[b]{.32\linewidth}
  \centering
  \centerline{\includegraphics[width=2.5cm]{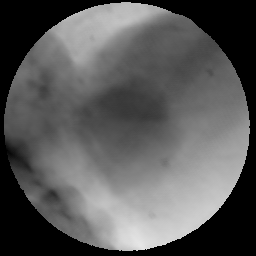}}
\end{minipage}
\begin{minipage}[b]{.32\linewidth}
  \centering
  \centerline{\includegraphics[width=2.5cm]{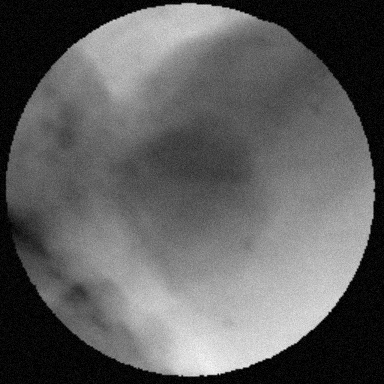}}
\end{minipage}
\begin{minipage}[b]{.32\linewidth}
  \centering
  \centerline{\includegraphics[width=2.5cm]{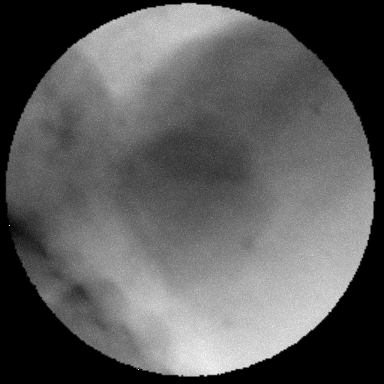}}
\end{minipage}
\caption{Noise-free DRR(1) and noised DRR with standard Gaussian noise(2) and our noise(3).}
\label{fig.8}
\end{figure}

\begin{figure}[H]
\begin{minipage}[b]{.45\linewidth}
  \centering
  \centerline{\includegraphics[width=4cm]{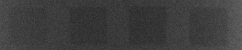}}
  \centerline{(a) Reference image}\medskip
\end{minipage}
\begin{minipage}[b]{.52\linewidth}
  \centering
  \centerline{\includegraphics[width=4cm]{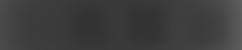}}
  \centerline{(b) Denoise with original SwinIR}\medskip
\end{minipage}
\begin{minipage}[b]{.45\linewidth}
  \centering
  \centerline{\includegraphics[width=4cm]{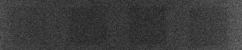}}
  \centerline{(b) Denoise with SwinIR-n}\medskip
\end{minipage}
\begin{minipage}[b]{.55\linewidth}
  \centering
  \centerline{\includegraphics[width=4cm]{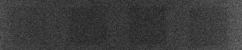}}
  \centerline{(b) Denoise with SwinIR-c}\medskip
\end{minipage}
\caption{These images are rows 161 to 211 and columns 70 to 312 of corresponding images. The SwinIR model with transfer learning achieves better denoising performance compared to when transfer learning is not applied. Without transfer learning, SwinIR blurs the boundaries of the aluminum blocks. Visually, the denoising effects of SwinIR-c and SwinIR-n are comparable.}
\label{fig.9}
\end{figure}

\section{Discussion}
\label{sec:typestyle}

The experimental results suggest that this approach captures noise characteristics more effectively than traditional Gaussian noise model. However, these findings are specific to the fixed geometry and parameters of the SyncTraX system, which may limit its generalizability to other IGRT configurations with different imaging conditions. Future research could investigate the robustness of the model across various equipment, providing insights for broader applications. Despite these limitations, this model establishes a critical foundation for enhancing IGRT image quality and improving tumor tracking in radiotherapy.

\section{Conclusion}
\label{sec:majhead}

This study introduces a statistical model of noise specifically designed for intraoperative SyncTraX fluoroscopic images used in IGRT, the results indicate that training SwinIR with noise images generated by our method achieved a 1.45 dB increase in PSNR and a 10\% improvement in VIF. The findings indicate that our approach more accurately replicates the noise patterns inherent to these images, thereby enhancing the denoising effectiveness of the SwinIR model. While our model is optimized for the configuration of SyncTraX system, future studies could adapt and test this approach across different IGRT systems to assess its generalizability and potential modifications for broader clinical use. Ultimately, this work contributes to advancing image quality in IGRT, offering a foundation for improving real-time tumor tracking accuracy.

\section{Acknowledgments}
\label{sec:acknowledgments}
 The usage of patients’ data was approved by the Institutional Review Board of the Yamaguchi University Hospital. We would like to express our gratitude to Dr. Yuki Yuasa, Mr. Wataru Mukaidani and Ms.Fu Weiwei, for their substantial assistance in this study.

\bibliographystyle{IEEEbib}
\bibliography{strings,refs}

\end{document}